\documentclass [12 pt]{article}
\def\be{\begin{equation}}
\def\eea{\end{eqnarray}}
\def\bea{\begin{eqnarray}}
\def\ee{\end{equation}}

\usepackage[psamsfonts]{amssymb}
\usepackage{amsmath}
\author{M. Amooshahi$^{1}$ \footnote{amooshahi-m@sci.sku.ac.ir} and F.
Kheirandish$^{2}$ \footnote{fardin$_{-}$
kh@phys.ui.ac.ir}
\\ $^{1}$ {\small Faculty of science, The University of Isfahan, Hezar Jarib Ave., Isfahan, Iran}
\\ $^{2}$ {\small  Faculty of science, The University of Isfahan, Hezar Jarib Ave., Isfahan,
Iran}}

\title{Electromagnetic field quantization in a magnetodielectric with external charges}
\begin{document}
\maketitle
\begin{abstract}

The electromagnetic field inside a cubic cavity filled up with a
linear magnetodielectric medium and in the presence of external
charges is quantized by modelling the magnetodielectric medium with
two independent quantum fields. Electric and magnetic polarization
densities of the medium are defined in terms of the ladder operators
of the medium and eigenmodes of the cavity. Maxwell
 and constitutive equations of the medium together with the equation of
motion of the charged particles have been obtained from the
Heisenberg equations using a minimal coupling scheme. Spontaneous
emission of a two level atom embedded in a magnetodielectric medium 
is calculated in terms of electric and magnetic susceptibilities of
the medium and the Green function of the cubic cavity as an
application of the model.


\end{abstract}
\section{Introduction}
The usual approach to quantum optics in non dispersive dielectric
media is to introduce the medium by its linear or nonlinear
susceptibilities. In one procedure, the macroscopic fields are
used to build an effective Lagrangian density whose Euler-Lagrange
equations are identical to the macroscopic Maxwell equations in
the dielectric, where the constitutive equations couple the
polarization field to the electric field \cite{1}-\cite{7}.
However, attempts to add dispersion to this effective scheme have
run into difficulties \cite{7,8}. The reason for this is that
inclusion of dispersion leads to a temporally nonlocal
relationship between the electric field and the displacement field
and the effective Lagrangian in this case is also nonlocal in time
and cannot be used directly in a quantization scheme. Another
problem is the inclusion of losses into the system. It is well
known that the dissipative nature of a medium is an immediate
consequence of its dispersive character and vice versa according
to the Kramers-Kronig relations\cite{9}. This suggests that in
order to quantize the electromagnetic field in a dielectric in a
way that is consistent with the Kramers-Kronig relations, one has
to introduce the medium into the formalism explicitly. This should
be done in such a away that the interaction between light and
matter will generate both dispersion and damping of the light
field. In an attempt to overcome these problems, by taking the
polarization of the medium as a dissipative quantum system and
based on the Hopfield model of a dielectric \cite{10,11}, a
canonical quantization of the electromagnetic field inside a
dispersive and absorptive dielectric can be presented, where the
polarization of the dielectric is modelled by a collection of
interacting matter fields\cite{12,13}. The absorptive character of
the medium is modelled through the interaction of the matter
fields with a reservoir consisting of a continuum of the
Klein-Gordon fields. In this model, eigen-operators for coupled
systems are calculated and the electromagnetic field has been
expressed in terms of them. The dielectric function is derived in
terms of a coupling function which couples the polarization of the
medium to the reservoir.

 One approach to study a quantum dissipative system tries to
 relate dissipation to the interaction between the system and a heat bath containing a collection
 of harmonic oscillators \cite{14}-\cite{26}.
 In this method, the whole system is composed of two parts, the main system and
 a bath which interacts with the main system and causes the dissipation of energy on it.
 In order to quantize the electromagnetic field in the presence of an absorptive magnetodielectric,
 it is reasonable to taking the electromagnetic field as the main quantum dissipative
system and the medium as the heat bath. In this point of view the
polarizability of the medium is a result of the properties of the
heat bath and accordingly, the polarizability should be defined in
terms of the dynamical variables of the medium. In this method,
contrary to the damped polarization model, polarizability and
absorptivity of the medium are not independent of each other
\cite{12,13}. Furthermore, if the medium is magnetizable, as well
as polarizable, the medium can be modelled with two independent
collection of harmonic oscillators, so that one of these
collections describes the electric properties and the other
describes the magnetic properties of the medium. This scheme
propose a consistent quantization of the electromagnetic field in
the presence of an absorptive magnetodielectric \cite{27}.

In the present paper, we generalize the quantization scheme
presented in \cite{27} to the case where there are some external
charges in the medium. We take a cubic cavity filled up with a
linear magnetodielectric, containing some external charges, as the
medium. As a simple application of the method, we study the
spontaneous decay of an initially excited two-level atom in a
dispersing and absorbing magnetodielectric.
\section{Quantum dynamics}
Quantum electrodynamics in a linear magnetodielectric medium, can
be accomplished by modelling the medium with two independent
quantum fields that interact with the electromagnetic field. One
of these quantum fields, namely E-quantum field, describes the
polarizability character of the medium and interacts with
displacement field through a minimal coupling term. The other
quantum field, M-quantum field, describes magnetizability
character of the medium and interacts with magnetic field through
a dipole interaction term. The Heisenberg equations for
electromagnetic field and the E- and M-quantum fields, give not
only the Maxwell equations but also the constitutive equations of
the medium and the equations of motion of external charges.

For this purpose let us consider an ideal rectangular cavity with
sides $L_1$, $L_2$ and $L_3$, along the coordinate axis $x$, $y$
and $z$, respectively, filled up with a magnetodielectric. The
vector potential of the electromagnetic field in coulomb gauge can
be expanded as \cite{28}
\begin{equation} \label{d1}
\vec{A}(\vec{r},t)=\sum_{\vec{n}}
\sum_{\lambda=1}^2\sqrt{\frac{4\hbar}
{\varepsilon_0V\omega_{\vec{n}}}}[a_{\vec{n}\lambda}(t)+a_{\vec{n}
\lambda}^\dag(t)]\vec{u}_{\vec{n}\lambda}(\vec{r}),
\end{equation}
where
\begin{eqnarray}\label{d1.1}
&& \vec{u}_{\vec{n}\lambda}(\vec{r})=e_x(\vec{n},\lambda)
f_1(\vec{n},\vec{r})\hat{x}+e_y(\vec{n},\lambda)
f_2(\vec{n},\vec{r})\hat{y}+e_z(\vec{n},\lambda)
f_3(\vec{n},\vec{r})\hat{z},\nonumber\\
&&f_1(\vec{n},\vec{r})=\cos\frac{n_1\pi x}{L_1}\sin\frac{n_2\pi
y}{L_2}\sin\frac{n_3\pi z}{L_3},\nonumber\\
&&f_2(\vec{n},\vec{r})=\sin\frac{n_1\pi x}{L_1}\cos\frac{n_2\pi
y}{L_2}\sin\frac{n_3\pi z}{L_3},\nonumber\\
&&f_3(\vec{n},\vec{r})=\sin\frac{n_1\pi x}{L_1}\sin\frac{n_2\pi
y}{L_2}\cos\frac{n_3\pi z}{L_3},
\end{eqnarray}
the mode vector $\vec{n}$ is a triplet of natural numbers $
(n_1,n_2,n_3) $, $ \sum_{\vec{n}}$ means $
\sum_{n_1,n_2,n_3=1}^\infty $, $ V=L_1L_2L_3 $ is the volume of
the cavity, $\varepsilon_0$ is the permittivity of the vacuum,
 $ \omega_{\vec{n}}=c\sqrt{\frac{n_1^2\pi^2}{L_1^2}+\frac{n_2^2\pi^2}
 {L_2^2}+\frac{n_3^2\pi^2}{L_3^2}}
$ is the frequency corresponding to the mode $\vec{n}$ and $
\vec{e}(\vec{n},\lambda), \hspace{00.30 cm} (\lambda=1,2) $ are
polarization vectors which satisfy
\begin{eqnarray}\label{d1.5}
\vec{e}(\vec{n},\lambda)\cdot\vec{e}(\vec{n},\lambda')&=&\delta_{\lambda\lambda'},\nonumber\\
 \vec{e}(\vec{n},\lambda)\cdot\vec{k}_{\vec{n}}&=&0,
\end{eqnarray}
where
$\vec{k}_{\vec{n}}=\frac{n_1\pi}{L_1}\vec{i}+\frac{n_2\pi}{L_2}\vec{j}+\frac{n_3\pi}{L_3}\vec{k}$
 is the wave  vector.
 Operators $a_{\vec{n}\lambda}(t)$ and $a_{\vec{n}\lambda}^\dag(t)
$ are annihilation and creation operators of the electromagnetic
field and satisfy the following equal time commutation rules
\begin{equation}\label{d2}
[a_{\vec{n}\lambda}(t),a_{\vec{m}\lambda'}^\dag(t)]=
\delta_{\vec{n},\vec{m}}\delta_{\lambda\lambda'}.
\end{equation}
Let $\vec{F}(\vec{r},t) $ be an arbitrary vector field, the
transverse component $ \vec{F}^\bot(\vec{r},t) $ and the
longitudinal component $ \vec{F}^\|(\vec{r},t) $ of $
\vec{F}(\vec{r},t) $ are defined as
\begin{eqnarray}\label{d3.3}
&& \vec{F}^\bot(\vec{r},t)=\vec{F}(\vec{r},t)+\int_V d^3r'
\nabla'\cdot\vec{F}(\vec{r'},t)\vec{\nabla} G(\vec{r},\vec{r'}), \\
&&\vec{F}^\|(\vec{r},t)=-\int_V d^3r'
\nabla'\cdot\vec{F}(\vec{r'},t)\vec{\nabla} G(\vec{r},\vec{r'}),
\end{eqnarray}
where $G(\vec{r},\vec{r'})$ is the Green function of the cubic
cavity
\begin{equation}\label{d3.4}
G(\vec{r},\vec{r'})=\frac{8}{V}\sum_{\vec{n}}
\frac{1}{|\vec{k}_{\vec{n}}|^2}\sin\frac{n_1\pi x
}{L_1}\sin\frac{n_2\pi y}{L_2}\sin\frac{n_3 \pi
z}{L_3}\sin\frac{n_1\pi x' }{L_1}\sin\frac{n_2\pi
y'}{L_2}\sin\frac{n_3 \pi z'}{L_3},
\end{equation}
In the presence of external charges the displacement field is not
purely transverse. The transverse component of the displacement
field can be expanded in terms of the cavity modes as
\begin{equation}\label{d3}
\vec{D}^\bot(\vec{r},t)=-i\varepsilon_0\sum_{\vec{n}}\sum_{\lambda=1}^2
\sqrt{\frac{4\omega_{\vec{n}}\hbar}{\varepsilon_0V}}
[a_{\vec{n}\lambda}^\dag(t)-a_{\vec{n}\lambda}(t)]\vec{u}_{\vec{n}\lambda}(\vec{r}).
\end{equation}
The commutation relations between the components of the vector
potential $\vec{A}$ and the transverse components of the
displacement field, $\vec{D}^\bot$, can be obtained from(\ref{d2})
as follows
\begin{equation}\label{d3.1}
[A_l(\vec{r},t),-D^\bot_j(\vec{r'},t)]=
\imath\hbar\delta_{lj}^\bot(\vec{r}-\vec{r'}),
\end{equation}
where $\delta_{lj}^\bot(\vec{r}-\vec{r'})$ is the transverse delta
function defined in terms of the eigenvector fields
$\vec{u}_{\vec{n}\lambda}$
\begin{equation}\label{d3.2}
\delta_{lj}^\bot(\vec{r}-\vec{r'})=\frac{8}{V}\sum_{\vec{n}}
\sum_{\lambda=1}^2u_{\vec{n}\lambda}^l(\vec{r})u_{\vec{n}\lambda}^j(\vec{r'})
=\frac{8}{V}\sum_{\vec{n}}(\delta_{lj}-\frac{k_{\vec{n}l}
k_{\vec{n}j}}{|\vec{k}_{\vec{n}}|^2})f_l(\vec{n},\vec{r})f_j(\vec{n},\vec{r'}).
\end{equation}
The Hamiltonian of the electromagnetic field inside a
magnetodielectric in the normal ordering form is
\begin{eqnarray}\label{d4.5}
H_F(t)&=&\int_V d^3r  : [\frac{ \vec{D}^\bot\cdot
\vec{D}^\bot}{2\varepsilon_0}+
\frac{(\nabla\times\vec{A})^2}{2\mu_0}] : ,\nonumber\\
&=&\sum_{\vec{n}}\sum_{\lambda=1}^2\hbar\omega_{\vec{n}}a_{\vec{n}\lambda}^\dag(t)
a_{\vec{n}\lambda}(t),
\end{eqnarray}
where $\mu_0$ is the magnetic permittivity of the vacuum and $
:\hspace{0.5 cm}:$ denotes the normal ordering operator. By
modelling the magnetodielectric medium with E- and M-quantum
fields we can write the Hamiltonian of the medium as the sum of
the Hamiltonians of the E- and M-quantum fields
\begin{eqnarray}\label{d4.6}
&&H_d=H_e+H_m\nonumber\\
&& H_e(t)=\sum_{\vec{n}}\sum_{\nu=1}^3\int_{-\infty}^{+\infty}d^3k
\hbar\omega_{\vec{k}}
d_{\vec{n}\nu}^\dag(\vec{k},t)d_{\vec{n}\nu}(\vec{k},t)
\nonumber\\
&&H_m(t)=\sum_{\vec{n}}\sum_{\nu=1}^3\int_{-\infty}^{+\infty}d^3k
\hbar\omega_{\vec{k}}
b_{\vec{n}\nu}^\dag(\vec{k},t)b_{\vec{n}\nu}(\vec{k},t) ,
\end{eqnarray}
where $ \omega_{\vec{k}}$ is dispersion relation of the medium.
Quantum Dynamics of a dissipative harmonic oscillator interacting
with an absorptive
 environment can
be investigated by modeling the environment with a continuum of
harmonic oscillators \cite{14}-\cite{18}. In the case of
quantization of electromagnetic field in the presence of a
magnetodielectric, electromagnetic field is the main dissipative
system and the medium play the role of the absorptive
environment. The Hamiltonian (\ref{d4.5})shows that
electromagnetic field contains a numerable set of harmonic
oscillators labeled by $\vec{n}$, $\lambda$. Therefore to each
harmonic oscillator of the electromagnetic field labeled by
$\vec{n}$, $\lambda$, a continuum of oscillators should be
corresponded. In the present scheme, to each harmonic oscillator
of the electromagnetic field labeled by $\vec{n}$, $\lambda$, we
have corresponded  two continuous set of harmonic oscillators
defined by the ladder operators $d_{\vec{n}\nu}(\vec{k}, t)$,
$d^{\dag}_{\vec{n}\nu}(\vec{k}, t)$ and $b_{\vec{n}\nu}(\vec{k},
t)$, $b^{\dag}_{\vec{n}\nu}(\vec{k}, t)$ which are to describe
the electric and magnetic properties of the medium respectively
and satisfy the equal time commutation relations
\begin{eqnarray}\label{d4.7}
&&[d_{\vec{n}\nu}(\vec{k},t),d_{\vec{m}\nu'}^\dag(\vec{k}',t)]=
\delta_{\vec{n},\vec{m}}\delta_{\nu\nu'}\delta(\vec{k}-\vec{k}'),\nonumber\\
&&[b_{\vec{n}\nu}(\vec{k},t),b_{\vec{m}\nu'}^\dag(\vec{k}',t)]=
\delta_{\vec{n},\vec{m}}\delta_{\nu\nu'}\delta(\vec{k}-\vec{k}').
\end{eqnarray}
If we want to extend the Huttner and Barnett model \cite{12} to a
magnetodielectric medium, we have to take a vector field $\vec{X}$
as electric polarization density of the medium and a heat bath $B$
interacting with $\vec{X}$ in order to take into account the
absorption due to the dispersion induced from the electrical
properties of the medium. Similarly, we must take a vector field
$\vec{Y}$ as magnetic polarization density and a heat bath
$\tilde{B}$( independent of $B$) interacting with $\vec{Y}$ in
order to take into account the absorption due to the dispersion
induced from the magnetic properties of the medium. Following the
Huttner and Barnett method, we find that a Fano diagonalization
\cite{11} process of the sum of the Hamiltonians related to the
electric polarization $\vec{X}$, the heat bath $B$ and their
interaction, lead to what we have named $H_e$ in Eq. (\ref{d4.6}).
Also a Fano diagonalization process of the sum of the Hamiltonians
related to the magnetic polarization $\vec{Y}$, the heat bath
$\tilde{B}$ and their interaction, lead to what we have named
$H_m$ in Eq.(\ref{d4.6}). If we use the Huttner and Barnett
approach inside a rectangular cavity, the labels $\vec{n}$ and
$\nu$ will appear in the expansion of the electric polarization
field $\vec{X}$ in terms of the eigenmodes of the cavity. On the
other hand, the heat bath $B$ interacting with the polarization
field $\vec{X}$ consists of a continuous set of harmonic
oscillators labelled by a continuous parameter $\omega$. It should
be noted that the three labels $\vec{n}$, $\nu$ and $\omega$
remain in $H_e$ after the diagonalization process.

 For a linear medium, the electric
polarization density operator can be written as a linear
combination of ladder operators $d_{\vec{n}\nu}(\vec{k},t)$ and
$d_{\vec{n}\nu}^\dag(\vec{k},t)$
\begin{equation}\label{d4.72}
\vec{P}(\vec{r},t)=\sqrt{\frac{8}{V}}\sum_{\vec{n}}\sum_{\nu=1}^3\int
d^3\vec{k}[f(\omega_{\vec{k}},\vec{r})d_{\vec{n}\nu}(\vec{k},t)+
h.c.] \vec{v}_{\vec{n}\nu}(\vec{r})
\end{equation}
with
\begin{eqnarray}\label{d4.725}
\vec{v}_{\vec{n}\nu}(\vec{r})&=&\vec{u}_{\vec{n}\nu}(\vec{r}),\hspace{0.5cm}\nu=1,2\nonumber\\
\vec{v}_{\vec{n}3}(\vec{r})&=&\hat{k}_{\vec{n}x}
f_1(\vec{n},\vec{r})\hat{x}+\hat{k}_{\vec{n}y}
f_2(\vec{n},\vec{r})\hat{y}+\hat{k}_{\vec{n}z}
f_3(\vec{n},\vec{r})\hat{z},\hspace{0.5cm}\hat{k}_{\vec{n}}=
\frac{\vec{k}_{\vec{n}}}{|\vec{k}_{\vec{n}}|}\nonumber\\
&&
\end{eqnarray}
the function $ f(\omega_{\vec{k}},\vec{r})$ is the coupling
function of electromanetic field and the E-quantum field. Also for
a linearly magnetizable medium we can express the magnetic
polarization density operator of the medium as a linear
combination of the ladder operators of the M-quantum
field\cite{27},
\begin{equation}\label{d4.73}
\vec{M}(\vec{r},t)=\sqrt{\frac{8}{V}}\sum_{\vec{n}}\sum_{\nu=1}^3\int
d^3\vec{k}[g(\omega_{\vec{k}},\vec{r})b_{\vec{n}\nu}(\vec{k},t)+h.c.]\vec{s}_{\vec{n}\nu}(\vec{r}).
\end{equation}
where the function $g(\omega_{\vec{k}},\vec{r})$ couples the
electromagnetic field to the M-quantum field and
\begin{eqnarray}\label{d4.735}
\vec{s}_{\vec{n}\nu}(\vec{r})&=&\frac{\nabla\times\vec{u}_{\vec{n}\nu}}
{|\vec{k}_{\vec{n}}|},\hspace{1cm}\nu=1,2,\nonumber\\
\vec{s}_{\vec{n}3}(\vec{r})&=&\hat{k}_{\vec{n}x}g_1(\vec{n},\vec{r})\hat{x}+
\hat{k}_{\vec{n}y}g_2(\vec{n},\vec{r})\hat{y}+\hat{k}_{\vec{n}z}g_3(\vec{n},\vec{r})\hat{z},\nonumber\\
g_1(\vec{n},\vec{r})&=&\sin\frac{n_1\pi x}{L_1}\cos\frac{n_2\pi
y}{L_2}\cos\frac{n_3\pi z}{L_3},\nonumber\\
g_2(\vec{n},\vec{r})&=&\cos\frac{n_1\pi x}{L_1}\sin\frac{n_2\pi
y}{L_2}\cos\frac{n_3\pi
z}{L_3},\nonumber\\
g_3(\vec{n},\vec{r})&=&\cos\frac{n_1\pi x}{L_1}\cos\frac{n_2\pi
y}{L_2}\sin\frac{n_3\pi z}{L_3}.
\end{eqnarray}
Now let the total Hamiltonian, i.e., electromagnetic field , the
E- and M-quantum field and the external charges be like this
\begin{eqnarray}\label{d4.55}
&&\tilde{H}(t)=\int_Vd^3r \left\{\frac{[
\vec{D}^\bot(\vec{r},t)-\vec{P}(\vec{r},t)]^2}{2\varepsilon_0}+
\frac{(\nabla\times\vec{A})^2(\vec{r},t)}{2\mu_0}
-\nabla\times\vec{A}(\vec(\vec{r},t)\cdot\vec{M}(\vec{r},t)\right\}+\nonumber\\
&&H_e+H_m+\sum_{\alpha=1}^N
\frac{[\vec{p}_\alpha-q_\alpha\vec{A}(\vec{r}_\alpha,t)]^2}{2m_\alpha}+\frac{1}{2\varepsilon_0}
\sum_{\alpha\neq\beta}q_\alpha
q_\beta G(\vec{r}_\alpha,\vec{r}_\beta)\nonumber\\
&&-\frac{1}{\varepsilon_0}\sum_{\alpha=1}^Nq_\alpha \int_V d^3r'
G(\vec{r}_\alpha,\vec{r'})\nabla'\cdot\vec{P}(\vec{r'},t)
\end{eqnarray}
where $q_\alpha $, $ m_\alpha $, $\vec{p}_\alpha $ and $
\vec{r}_\alpha $ are
  charge, mass, linear momentum and position of the $\alpha$th particle respectively.
  The function $G$ is the Green function given by (\ref{d3.4}).
Using (\ref{d3.1}), the Heisenberg equation for $\vec{A}$ and $
\vec{D}^\bot $ can be obtained as
\begin{equation}\label{d5.1}
\frac{\partial\vec{A}(\vec{r},t)}{\partial
t}=\frac{\imath}{\hbar}[\tilde{H},\vec{A}(\vec{r},t)]=
-\frac{\vec{D}^\bot(\vec{r},t)-\vec{P}^\bot(\vec{r},t)}{\varepsilon_0},
\end{equation}
\begin{equation}\label{d6.1}
\frac{\partial\vec{D}^\bot(\vec{r},t)}{\partial
t}=\frac{\imath}{\hbar}[\tilde{H},\vec{D}^\bot(\vec{r},t)]=
\frac{\nabla\times\nabla\times\vec{A}(\vec{r},t)}{\mu_0}-\nabla\times\vec{M}^\bot(\vec{r},t)-\vec{J}^\bot(\vec{r},t),
\end{equation}
where
\begin{equation}\label{d7}
J_i^\bot(\vec{r},t)=\sum_{\alpha=1}^N\sum_{l=1}^3 q_\alpha
\frac{d\vec{r}_\alpha}{dt}\delta_{li}^\bot(\vec{r}_\alpha(t)-\vec{r})
\end{equation}
is the transverse component of external current density. If we
define transverse electrical field $\vec{E}^\bot $, induction $
\vec{B}$ and magnetic field $\vec{H}$ as
 \begin{equation}\label{d8}
 \vec{E}^\bot=-\frac{\partial\vec{A}}{\partial t},\hspace{1.00
 cm}\vec{B}=\nabla\times\vec{A},\hspace{1.00
 cm}\vec{H}=\frac{\vec{B}}{\mu_0}-\vec{M},
 \end{equation}
 then (\ref{d5.1}) and (\ref{d6.1}) can be rewritten as
\begin{equation} \label{d9}
\vec{D}^\bot=\varepsilon_0 \vec{E}^\bot+\vec{P}^\bot,
\end{equation}
\begin{equation}\label{d10}
\frac{\partial \vec{D}^\bot}{\partial
t}+\vec{J}^\bot=\nabla\times\vec{H},
\end{equation}
as expected. In the presence of external charges, the longitudinal
components of the electrical and displacement field can be written
as
\begin{equation}\label{d10.1}
\vec{E}^\|(\vec{r},t)=-\frac{1}{\varepsilon_0}\sum_{\alpha=1}^N
q_\alpha
\vec{\nabla}_rG(\vec{r},\vec{r}_\alpha(t))-\frac{\vec{P}^\|}{\varepsilon_0},
\end{equation}
\begin{equation}\label{d10.2}
\vec{D}^\|(\vec{r},t)=\varepsilon_0\vec{E}^\|+\vec{P}^\|=-\sum_{\alpha=1}^N
q_\alpha\vec{\nabla}_rG(\vec{r},\vec{r}_\alpha(t)),
\end{equation}
where $\vec{P}^\|$ is  the longitudinal component of electric
polarization density. If we apply the Heisenberg equation to the
ladder operators of the medium, we obtain the macroscopic
constitutive equations of the medium which relate the electric and
magnetic polarization densities to electric and magnetic field,
respectively. Using commutation relations (\ref{d4.7}), we find
from Heisenberg equations
\begin{eqnarray}\label{d10.3}
&&\dot{d}_{\vec{n}\nu}(\vec{k},t)=
\frac{\imath}{\hbar}[\tilde{H},d_{\vec{n}\nu}(\vec{k},t)]=\nonumber\\
&&-\imath\omega_{\vec{k}}d_{\vec{n}\nu}(\vec{k},t)+\frac{\imath}{\hbar}
\sqrt{\frac{8}{V}}\int_V d^3\vec{r'}f^*(\omega_{\vec{k}},\vec{r'})
\vec{E}(\vec{r'},t)\cdot\vec{v}_{\vec{n}\nu}(\vec{r'}),
\end{eqnarray}
\begin{eqnarray}\label{d11}
&&\dot{b}_{\vec{n}\nu}(\vec{k},t)=\frac{\imath}{\hbar}[\tilde{H},b_{\vec{n}\nu}(\vec{k},t)]=
\nonumber\\
&-&\imath\omega_{\vec{k}}b_{\vec{n}\nu}(\vec{k},t)
+\frac{\imath}{\hbar}\sqrt{\frac{8}{V}} \int_V
d^3r'g^*(\omega_{\vec{k}},\vec{r'})\vec{B}(\vec{r'},t)\cdot\vec{s}_{\vec{n}\nu}(\vec{r'}),
\end{eqnarray}
with the following formal solutions
\begin{eqnarray}\label{d11.1}
&&{d}_{\vec{n}\nu}(\vec{k},t)=
d_{\vec{n}\nu}(\vec{k},0)e^{-\imath\omega_{\vec{k}}t}+\nonumber\\
&&\frac{\imath}{\hbar}\sqrt{\frac{8}{V}}\int_0^t
dt'e^{-\imath\omega_{\vec{k}}(t-t')} \int_V
d^3r'f^*(\omega_{\vec{k}},\vec{r'})
\vec{E}(\vec{r'},t')\cdot\vec{v}_{\vec{n}\nu}(\vec{r'}),
\end{eqnarray}
\begin{eqnarray}\label{d11.2}
&&{b}_{\vec{n}\nu}(\vec{k},t)=
b_{\vec{n}\nu}(\vec{k},0)e^{-\imath\omega_{\vec{k}}t}+\nonumber\\
&&\frac{\imath}{\hbar}\sqrt{\frac{8}{V}}\int_0^t
dt'e^{-\imath\omega_{\vec{k}}(t-t')} \int_V
d^3r'g^*(\omega_{\vec{k}},\vec{r'})\vec{B}
(\vec{r'},t')\cdot\vec{s}_{\vec{n}\nu}(\vec{r'}).
\end{eqnarray}
 By substituting (\ref{d11.1}) in (\ref{d4.72}) and using the completeness relations
\begin{equation}\label{d11.3}
\sum_{\vec{m}}\sum_{\nu=1}^{3}v_{\vec{m}\nu}^{\alpha}(\vec{r})
v_{\vec{m}\nu}^{\beta}(\vec{r'})=\sum_{\vec{m}}\sum_{\nu=1}^{3}s_{\vec{m}\nu}^{\alpha}
(\vec{r})s_{\vec{m}\nu}^{\beta}(\vec{r'})=\frac{V}{8}\delta_{\alpha\beta}\delta
(\vec{r}-\vec{r'}),
\end{equation}
we can find the constitutive equation
\begin{equation}\label{d12}
\vec{P}(\vec{r},t)=\vec{P}_N(\vec{r},t)+\varepsilon_0\int_0^{|t|}
d t' \chi_e(\vec{r},|t|-t')\vec{E}(\vec{r},\pm t'),
\end{equation}
for the medium which relates the electric polarization density to
the total electric field, where the upper (lower) sign,
corresponds to $t>0$ ($ t<0 $) respectively, and
$\vec{E}=\vec{E}^\bot+\vec{E}^\|$ is the total electrical field
with $\vec{E}^\bot$ and $ \vec{E}^\|$ defined in (\ref{d8}) and
(\ref{d10.1}). The memory function
\begin{equation}\label{d12.1}
\chi_e(\vec{r},t)=\left\{\begin{array}{cc}
 \frac{8\pi}{\hbar \varepsilon_0}\int_0^\infty d |\vec{k}|
|\vec{k}|^2 |f(\omega_{\vec{k}},\vec{r})|^2 \sin\omega_{\vec{k}}t
& \hspace{1.00cm}t>0\\
\\
 0 & \hspace{1.00cm}t\leq 0
\end{array}\right.
\end{equation}
is the electric susceptibility and in frequency domain satisfies
the following relations
\begin{equation}\label{d12.2}
Im[\underline{\chi}_e(\vec{r},\omega)=
\frac{4\pi^2}{\hbar\varepsilon_0}|f(\omega,
\vec{r})|^2\sum_{i}\frac{|\vec{k}_i|^2}{|\frac{d\omega}{d|\vec{k}|}(|\vec{k}|=
|\vec{k}_i|)|},
\end{equation}
\begin{equation}\label{d12.3}
Re[\underline{\chi}_e(\vec{r},\omega)=
\frac{8\pi}{\hbar\varepsilon_0}\int_{0}^{\infty}d |\vec{k}|
|\vec{k}|^2
|f(\omega_{\vec{k}},\vec{r})|^2\frac{\omega_{\vec{k}}}{\omega_{\vec{k}}^2-\omega^2}
\end{equation}
where
\begin{equation} \label{d12.4}
\underline{\chi}_e(\vec{r},\omega)=\int_{0}^{\infty} d t
\chi_e(\vec{r},t) e^{\imath \omega t}
\end{equation}
is the Fourier transform of the electric susceptibility and
$|\vec{k}_i|$s are the roots of the algebraic equation
$\omega=\omega_{\vec{k}}\equiv\omega(|\vec{k}|)$. A feature of the
present quantization method is it's flexibility in choosing a
dispersion relation and a coupling function such that they satisfy
the  relation (\ref{d12.2}). In other words, the dispersion
relation and the coupling function $f(\omega_{\vec{k}},\vec{r})$
are two free parameters of this theory up to the constraint
relation (\ref{d12.2}). Various choices of $\omega_{\vec{k}}$ and
$f(\omega_{\vec{k}}, \vec{r})$ satisfying (\ref{d12.2}) do not
change the commutation relations between electromagnetic field
operators and lead to equivalent expressions for the field
operators and physical observables. As can be seen from the
relation (\ref{d12.2}), in order to have a finite value for
susceptibility for any frequency $\omega$,we should assume that
the denominator of the fraction in (\ref{d12.2}) is non zero.  so
we assume that the dispersion relation is a monotonic function of
$|\vec{k}|$ . For simplicity and simple calculations we take a
linear dispersion relation $\omega = c|\vec{k}|$ which leads to
the following relation for the electric susceptibility:\\
\begin{equation}\label{d13}
\chi_e(\vec{r},t)=\left\{\begin{array}{cc}
 \frac{8\pi}{\hbar c^3 \varepsilon_0}\int_0^\infty d\omega
\omega^2|f(\omega ,\vec{r})|^2\sin\omega t  & \hspace{1cm}t>0 \\
\\
0 & \hspace{1cm}t\leq 0
\end{array}\right.
\end{equation}
 The operator $\vec{P}_N(\vec{r},t) $ in
(\ref{d12}), is the noise electric polarization density
\begin{equation}\label{d14}
\vec{P}_N(\vec{r},t)=
\sqrt{\frac{8}{V}}\sum_{\vec{n}}\sum_{\nu=1}^3\int
d^3\vec{k}[f(\omega_{\vec{k}},\vec{r})d_{\vec{n}\nu}(\vec{k},0)
e^{-\imath\omega_{\vec{k}}t}+h.c.]\vec{v}_{\vec{n} \nu}(\vec{r}).
\end{equation}
Similarly by substituting (\ref{d11.2}) in (\ref{d4.73}), we
obtain the following expression for $\vec{M}(\vec{r},t)$
\begin{equation}\label{d15}
\vec{M}(\vec{r},t)=\vec{M}_N(\vec{r},t)+\frac{1}{\mu_0}\int_0^{|t|}
dt' \chi_m(\vec{r},|t|-t')\vec{B}(\vec{r},\pm t'),
\end{equation}
where $\chi_m $ is the magnetic susceptibility of the medium
\begin{eqnarray}\label{d16}
\chi_m(\vec{r},t)&=&\frac{8\pi\mu_0}{\hbar c^3}\int_0^\infty
d\omega\omega^2|g(\omega,\vec{r})|^2
\sin\omega t,\hspace{1cm}t>0,\nonumber\\
\chi_m(\vec{r},t)&=&0,\hspace{6.25cm}t\leq 0.
\end{eqnarray}
and operator $M_N(\vec{r},t)$ is the noise magnetic polarization
density
\begin{equation}\label{d17}
\vec{M}_N(\vec{r},t)=
\sqrt{\frac{8}{V}}\sum_{\vec{n}}\sum_{\nu=1}^3\int
d^3\vec{k}[g(\omega_{\vec{k}},\vec{r})b_{\vec{n}\nu}(\vec{k},0)
e^{-\imath\omega_{\vec{k}}t}+H.C] \vec{s}_{\vec{n}\nu}(\vec{r}).
\end{equation}
If we are given a definite $\chi_e(t)$ and $\chi_m(t)$, then we
can inverse the relations (\ref{d13}) and (\ref{d19}) and obtain
the corresponding coupling functions $f$ and $g$ as
\begin{eqnarray}\label{d18}
|f(\omega,\vec{r})|^2&=&\frac{\hbar c^3\varepsilon_0
}{4\pi^2\omega^2}\int_0^\infty dt\chi_e(\vec{r},t)
\sin\omega t,\hspace{1cm}\omega>0,\nonumber\\
|f(\omega,\vec{r})|^2&=&0,\hspace{5.50cm}\omega=0.
\end{eqnarray}
\begin{eqnarray}\label{d19}
|g(\omega,\vec{r})|^2&=&\frac{\hbar c^3
}{4\pi^2\mu_0\omega^2}\int_0^\infty dt\chi_m(\vec{r},t)
\sin\omega t,\hspace{1cm}\omega>0,\nonumber\\
|g(\omega,\vec{r})|^2&=&0,\hspace{6.00cm}\omega=0.
\end{eqnarray}

Therefore the constitutive equations (\ref{d9}), (\ref{d12}) and
(\ref{d17}) together with the Maxwell equations, can be obtained
directly from the Heisenberg equations applied to the
electromagnetic field and the quantum fields
$ E $ and $ M $.\\
It is clear from (\ref{d14}) and (\ref{d17}) that the explicit
forms of the noise polarization densities are known. Also because
the coupling functions $ f,g $ are common factors in the noise
densities $ \vec{P}_N  , \vec{M}_N $, and susceptibilities $
\chi_e ,\chi_m $, it is clear that the strength of the noise
fields are dependent on the strength of $\chi_e ,\chi_m $, which
describe the dissipative character of a magnetodielectric medium.
Specially when the medium tends to a non absorbing one, the noise
polarizations tend to zero and this quantization scheme is reduced
to the usual quantization in this medium \cite{27}.

Finally  using the total Hamiltoian (\ref{d4.55}) and the
commutation relations
\begin{equation}\label{d32.9}
[\vec{r}_\alpha , \vec{p}_\beta]=\imath\hbar\delta_{\alpha\beta}I
\end{equation}
 it is easy to show that the Heisenberg equations of motion for charged
particles are
\begin{eqnarray}\label{d33}
&&m_\alpha\ddot{\vec{r}}_\alpha=q_\alpha\vec{E}(\vec{r}_\alpha,t)+\nonumber\\
&&\frac{q_\alpha}{2}
[\dot{\vec{r}}_\alpha\times\vec{B}(\vec{r}_\alpha,t)+\vec{B}(\vec{r}_\alpha,t)
\times\dot{\vec{r}}_\alpha]\hspace{1.00cm}\alpha=1,2,...,N
\end{eqnarray}
\section{Spontaneous emission of an excited two-level atom in the presence
of a magnetodielectric medium}
 In this section as a simple application we use the
quantization scheme in the previous section to calculate the
spontaneous decay rate of an initially excited two-level atom
embedded in a magnetodielectric. For this purpose, let us consider
a one electron atom with position $\vec{R}$ which interacts with
the electromagnetic field in the presence of a magnetodielectric
medium. If we restrict our attention to the electric-dipole
approximation\cite{29,30}, then the Hamiltonian (\ref{d4.55}) can
be approximated as
\begin{eqnarray}\label{d34}
&&\tilde{H}=H_0+H',\nonumber\\
&&H_0=H_F+H_e+H_m+H_{at},\nonumber\\
&&H'=\int_{V}d^3r'\left[-\frac{\vec{D}^\bot\cdot
\vec{P}}{\varepsilon_0}-\vec{\nabla}\times \vec{A}\cdot
\vec{M}\right]-e\vec{r}\cdot
\vec{E}_D(\vec{R},t),\nonumber\\
&&
\end{eqnarray}
where
\begin{eqnarray}\label{d35}
&&H_{at}=\frac{\vec{p}^2}{2m}+\frac{e}{\varepsilon_0}\sum_{q_\alpha\neq
e}q_\alpha G(\vec{r},\vec{r}_\alpha),\nonumber\\
&&\vec{E}_D(\vec{R},t)=\vec{E}^\bot(\vec{R},t)-\frac{\vec{P}^\|(\vec{R},t)}{\varepsilon_0}
=\frac{\vec{D}^\bot(\vec{R},t)}{\varepsilon_0}-\frac{\vec{P}(\vec{R},t)}{\varepsilon_0},
\end{eqnarray}
and $e$, $m$, $\vec{r}$ and $\vec{p}$ are charge, mass, position
and momentum of the atom electron. In Hamiltonian (\ref{d34}) we
have ignored from $ \frac{e^2}{2m}\vec{A}^2(\vec{R},t)$ and $
\frac{1}{2\varepsilon_0}\int_{V}d^3r \vec{P}^2(\vec{r},t) $ ,
because these terms do not affect the decay rate or level shifts
in the dipole approximation and therefore will not be of concern
here. For reality we assume that the atom is localized in a free
space region $V_0$ without polarization density. Then for a
two-level atom with upper state $|2\rangle$, lower state
$|1\rangle$ and transition frequency $\omega_0$, the Hamiltonian
(\ref{d34}) can be written as \cite{28}
\begin{eqnarray}\label{d36}
&&\tilde{H}=\hbar\omega_0\sigma^\dag\sigma+H_F+H_e+H_m+H_{at}\nonumber\\
&&+\int_{\Omega}d^3r'\left[-\frac{\vec{D}^\bot\cdot
\vec{P}}{\varepsilon_0}-\vec{\nabla}\times \vec{A}\cdot
\vec{M}\right]-\frac{e}{\varepsilon_0}(\vec{r}_{12}\sigma+\vec{r}^*_{12}\sigma^\dag),
\cdot\vec{D}^\bot(\vec{R},t)\nonumber\\
&&
\end{eqnarray}
 where $\Omega=V-V_0$, is the region containing the magnetodielctric, $\sigma=|1\rangle\langle 2|$ and
$\sigma^\dag=|2\rangle\langle 1|$ are Pauli operators of the
two-level atom and $ \vec{r}_{12}=\langle1|\vec{r}|2\rangle$ is
its transition dipole momentum. It is remarkable that the
expansions (\ref{d4.72} ) and (\ref{d4.73}) for electric and
magnetic polarization densities are independent of the shape of
the magnetodielectric and depend only on the shape of the cavity
contained the magnetodielectric. The shape of the medium affects
the coupling functions $f(\omega,\vec{r})$ and
$g(\omega,\vec{r})$ and the time dependence of the ladder
operators of the medium, such that for a given medium with
definite susceptibilities $\chi_e$ and $\chi_m$, the dynamics of
the total system leads to the correct constitutive equations
(\ref{d12}) and(\ref{d15}) where the coupling functions are given
by (\ref{d18}) and (\ref{d19}). This is just as the expansion
(\ref{d1}) for the vector potential of the vacuum field where the
presence or absence of the external charges do not change the
form of the expansion and the presence of the external charges
merely affect the time dependence of the ladder operators
$a_{\vec{n},\lambda}$, $a^{\dag}_{\vec{n},\lambda}$.\\
To study the
spontaneous decay of an initially excited atom we may look for
the wave function
 of the total system in the framework of the Weisskopf-Wigner
 theory\cite{29,31},
\begin{eqnarray}\label{d37}
 |\psi(t)\rangle &=&c(t)|2\rangle|0\rangle_F|0\rangle_e|0\rangle_m+
\sum_{\vec{n}}\sum_{\lambda=1}^2
F_{\vec{n}\lambda}(t)|1\rangle|\vec{n},\lambda\rangle_F|0\rangle_e|0\rangle_m
\nonumber\\
&+&\sum_{\vec{m}}\sum_{\nu=1}^3\int d^3\vec{k}
D_{\vec{m}\nu}(\vec{k},t)|1\rangle|0\rangle_F|
\vec{m},\vec{k},\nu\rangle_e|0\rangle_m\nonumber\\
&+&\sum_{\vec{m}}\sum_{\nu=1}^3\int d^3\vec{k}
M_{\vec{m}\nu}(\vec{k},t)|1\rangle|0\rangle_F|0
\rangle_e|\vec{m},\vec{k},\nu\rangle_m,
\end{eqnarray}
 where
$|0\rangle_F$, $|0\rangle_e$ and $|0\rangle_m $ are vacuum state
of electromagnetic field and E- and M-quantum fields,
respectively. The coefficients $ c(t) $, $ F_{\vec{n}\lambda}(t)
$, $ D_{\vec{m}\nu}(\vec{k},t) $ and $ M_{\vec{m}\nu}(\vec{k},t) $
are to be specified by Schr\"{o}dinger equation
\begin{equation}\label{d38}
i\hbar\frac{\partial |\psi(t)\rangle}{\partial
t}=H|\psi(t)\rangle,
\end{equation}
for initial conditions $ c(0)=1 ,  F_{\vec{n}\lambda}(0)=
D_{\vec{m}\nu}(\vec{k},0) = M_{\vec{m}\nu}(\vec{k},0)=0 $.
Substituting $ |\psi(t)\rangle $ from (\ref{d37}) in (\ref{d38})
and using the expansions  (\ref{d3}), (\ref{d4.72}), (\ref{d4.73})
and applying the rotating wave approximation(RWA) \cite{29,30}, we
find the following coupled differential equations for the
coefficients of the wave function (\ref{d37})
\begin{equation}\label{d39a}
i\hbar\dot{M}_{\vec{m}\nu}(\vec{k},t)=\hbar\omega_{\vec{k}}M_{\vec{m}\nu}(\vec{k},t)-\sum_{\vec{n}}\sum_{\lambda=1}^2
[L_{\vec{n}\lambda}^{\vec{m}\nu}(\omega_{\vec{k}})]^*F_{\vec{n}\lambda}(t),
\end{equation}
\begin{eqnarray}\label{d39b}
&&i\hbar\dot{D}_{\vec{m}\nu}(\vec{k},t)=\hbar\omega_{\vec{k}}D_{\vec{m}\nu}(\vec{k},t)+\sum_{\vec{n}}\sum_{\lambda=1}^2
[Q_{\vec{n}\lambda}^{\vec{m}\nu}(\omega_{\vec{k}})]^*F_{\vec{n}\lambda}(t)+\nonumber\\
&&
\end{eqnarray}
\begin{eqnarray}\label{d39c}
&&i\hbar\dot{F}_{\vec{n}\lambda}(t)=\hbar\omega_{\vec{n}}F_{\vec{n}\lambda}(t)-\sum_{\vec{m}}\sum_{\nu=1}^3\int
d^3\vec{k} L_{\vec{n}\lambda}^{\vec{m}\nu}(\omega_{\vec{k}})
M_{\vec{m}\nu}(\vec{k},t)+\nonumber\\
&&\sum_{\vec{m}}\sum_{\nu=1}^3\int d^3\vec{k}
Q_{\vec{n}\lambda}^{\vec{m}\nu}(\omega_{\vec{k}})D_{\vec{m}\nu}(\vec{k},t)
+\left[ie\sqrt{\frac{4\hbar\omega_{\vec{n}}}{\varepsilon_0 V}}\vec{r}_{12}\cdot \vec{u}_{\vec{n}\lambda}
(\vec{R})\right]c(t),\nonumber\\
&&
\end{eqnarray}
\begin{eqnarray}\label{d39d}
&&i\hbar\dot{c}(t)=\hbar\omega_0c(t)-\left[ie\sum_{\vec{n}}\sum_{\lambda=1}^2
\sqrt{\frac{4\hbar\omega_{\vec{n}}}{\varepsilon_0 V}}\vec{r}^*_{12}\cdot
\vec{u}_{\vec{n}\lambda}(\vec{R})\right]F_{\vec{n}\lambda}(t),\nonumber\\
&&
\end{eqnarray}
 where
\begin{eqnarray}\label{d40}
&&Q_{\vec{n}\lambda}^{\vec{m}\nu}(\omega_{\vec{k}})=i\sqrt{\frac{32\hbar
\omega_{\vec{n}}}{\varepsilon_0 V^2}}
\int_{V-V_0}d^3r' f(\omega_{\vec{k}},\vec{r'})\vec{u}_{\vec{n}\lambda}(\vec{r'})\cdot
\vec{v}_{\vec{m}\nu}(\vec{r'}),\nonumber\\
&&L_{\vec{n}\lambda}^{\vec{m}\nu}(\omega_{\vec{k}})=\sqrt{\frac{32\hbar
\mu_0\omega_{\vec{n}}}{ V^2}}
\int_{V-V_0}d^3r' g(\omega_{\vec{k}},\vec{r'})\vec{s}_{\vec{n}\lambda}(\vec{r'})\cdot
\vec{s}_{\vec{m}\nu}(\vec{r'}).\nonumber\\
 &&
\end{eqnarray}
One can solve these coupled differential equations  by Laplace
transformation. Let $ \tilde{f}(\rho)$ denotes the Laplace
transformation of $ f(t)$, then taking the the Laplace
transformation of equations (\ref{d39a})-(\ref{d39c}) and then
their combination and using the completeness relations
(\ref{d11.3}) we find
\begin{eqnarray}\label{d42}
&&\left[\imath\hbar\rho-\hbar\omega_{\vec{n}}\right]\tilde{F}_{\vec{n}\lambda}(\rho)=\hbar\omega_{\vec{n}}\sum_{\vec{n'}}\sum_{\lambda'=1}^2
W_{\vec{n}\lambda}^{\vec{n'}\lambda'}\tilde{F}_{\vec{n'}\lambda'}(\rho)+
\left[\imath e\sqrt{\frac{4\hbar\omega_{\vec{n}}}{\varepsilon_0
V}}\vec{r}_{12}\cdot\vec{u}_{\vec{n}\lambda}(\vec{R})\right]\tilde{c}(\rho),\nonumber\\
&&
\end{eqnarray}
where we have applied the initial conditions $
F_{\vec{n}\lambda}(0)=D_{\vec{m}\nu}(\vec{k},0)=
M_{\vec{m}\nu}(\vec{k},0)=0 $ and
\begin{eqnarray}\label{d43}
&&W_{\vec{n}\lambda}^{\vec{n'}\lambda'}(\imath\rho)=\frac{8}{V}\sqrt{\frac{\omega_{\vec{n'}}}
{\omega_{\vec{n}}}}\times\nonumber\\
&& \left\{ \int_{\Omega}d^3r\left[
Z_e(\imath\rho,\vec{r})\vec{u}_{\vec{n}\lambda}(\vec{r})\cdot
\vec{u}_{\vec{n'}\lambda'}(\vec{r})+
 Z_m(\imath\rho,\vec{r})\vec{s}_{\vec{n}\lambda}(\vec{r})\cdot
 \vec{s}_{\vec{n'}\lambda'}(\vec{r})\right]\right\},\nonumber\\
 &&
 \end{eqnarray}
\begin{eqnarray}\label{d44}
&&Z_e(\imath\rho,\vec{r})=\frac{1}{2\pi}\int_0^\infty
d\omega\frac{\chi_{ei}(\omega,\vec{r})}{\imath\rho-\omega},\hspace{1.50cm}Z_m(\imath\rho,\vec{r})=
\frac{1}{2\pi}\int_0^\infty
d\omega\frac{\chi_{mi}(\omega,\vec{r})}{\imath\rho-\omega}.\nonumber\\
&&
\end{eqnarray}
Here $\chi_{ei}(\omega,\vec{r})$ and $\chi_{mi}(\omega,\vec{r})$
are the imaginary parts of electric and magnetic susceptibilities
in frequency domain, respectively. Equation (\ref{d42}) is a
complicated algebraic equation for $
\tilde{F}_{\vec{n}\lambda}(\rho) $ and may be solved by iteration
method. In the first step of an iteration method
$\tilde{F}_{\vec{n}\lambda}$ may be approximated by
\begin{equation}\label{d45}
\tilde{F}_{\vec{n}\lambda}(\rho)=\tilde{F}_{\vec{n}\lambda}^{(0)}(\rho)+
\sum_{\vec{n'}}\sum_{
\lambda'=1}^2\frac{\omega_{\vec{n}}W_{\vec{n}\lambda}^{\vec{n'}\lambda'}
(\imath\rho)\tilde{F}_{\vec{n'}\lambda'}^{(0)}( \rho)
}{\imath\rho-\omega_{\vec{n}}[1+W_{\vec{n}\lambda}^{\vec{n}\lambda}(\imath\rho)]},
\end{equation}
where
\begin{equation}\label{d46}
\tilde{F}_{\vec{n}\lambda}^{(0)}(\rho)=
 \frac{\imath e
\sqrt{\frac{4\hbar\omega_{\vec{n}}}{\varepsilon_0 V
}}\vec{r}_{12}\cdot\vec{u}_{\vec{n}\lambda}(\vec{R})}{\imath\hbar\rho-\hbar\omega_{\vec{n}}}\tilde{c}(\rho),
\end{equation}
is the solution of (\ref{d42}) in the absence of medium, that is
when $W_{\vec{n}\lambda}^{\vec{n'}\lambda'}=0$. Now combination of
equation (\ref{d45}) and the Laplace transformations of the
equation (\ref{d39d}) yields
\begin{eqnarray}\label{d47}
\imath\hbar(\tilde{c}(\rho)-c(0))=\hbar\omega_0
\tilde{c}(\rho)+\left[\vec{r}_{12}^*\cdot\left(\tilde{G}^{(0)}(\vec{R},\vec{R},\imath\rho)+\tilde{G}
(\vec{R},\vec{R},\imath\rho)\right)\cdot\vec{r}_{12}\right]\tilde{c}(
\rho),\nonumber\\
&&
\end{eqnarray}
where
\begin{equation}\label{d48}
 \tilde{G}^{(0)}(\vec{R},\vec{R},\imath\rho)=\frac{4
e^2}{\varepsilon_0
V}\sum_{\vec{n}}\sum_{\lambda=1}^2\frac{\omega_{\vec{n}}\vec{u}_{\vec{n}\lambda}(\vec{R})
\vec{u}_{\vec{n}\lambda}(\vec{R})}{ \imath\rho-\omega_{\vec{n}}},
\end{equation}
\begin{eqnarray}\label{d49}
&& \tilde{G}(\vec{R},\vec{R},\imath\rho)=\nonumber\\
&&\frac{4 e^2}{\varepsilon_0
V}\sum_{\vec{n},\vec{n'}}\sum_{\lambda,\lambda'=1}^2
\frac{\sqrt{\omega_{\vec{n}}^3\omega_{\vec{n'}}}\vec{u}_{\vec{n}\lambda}(\vec{R})W_{\vec{n}\lambda}^{\vec{n'}\lambda'}(
\imath\rho)\vec{u}_{\vec{n'}\lambda'}(\vec{R})}{\left[
\imath\rho-\omega_{\vec{n}}\left(1+W_{\vec{n}\lambda}^{\vec{n}\lambda}(\imath\rho)\right)\right]
\left[\imath\rho-\omega_{\vec{n'}}\right]}.\nonumber\\
&&
\end{eqnarray}
The diadic $\tilde{G}^{(0)}$ gives the contribution of the vacuum
field for the spontaneous emission and level shift of the atom,
that is, when there is no magnetodielectric, while the diadic
$\tilde{G}$ gives the spontaneous and frequency shift of the atom
due to the presence of the medium.
 From definitions (\ref{d43}) and (\ref{d44}) it is obvious that $
Z_e(\imath\rho,\vec{r})$, $ Z_e(\imath\rho,\vec{r})$ and $
W_{\vec{n}\lambda}^{\vec{n'}\lambda'}(\imath\rho) $ are analytic
functions in complex semi plane $ Re(\rho)>0$ and since the
denominator in diadic $\tilde{G}$ has no zero in semi plane $
Re(\rho)>0$ so $\tilde{G}(\vec{R},\vec{R},\imath\rho)$ is analytic
for $ Re(\rho)>0$. Taking the inverse Laplace transformation of
(\ref{d46}) we find the following integro-differential equation
\begin{equation}\label{d50}
\dot{c}(t)=-\imath\omega_0c(t)-(\Gamma_0+\imath\Delta_0)c(t)+\int_0^t
K(t-t')c(t')dt',
\end{equation}
 where
\begin{equation}\label{d51}
K(t-t')= \frac{1}{2\pi}\int_{-\infty}^{+\infty} d \omega
e^{-\imath\omega(t-t')}\left[\vec{r}_{12}\cdot\left(\tilde{G}(\vec{R},\vec{R},\omega+\imath\tau)
\right)\cdot\vec{r}_{12}\right],
\end{equation}
and $ \Gamma_0 $ and $\Delta_0$ are contribution of the vacuum
field for the decay rate and the lamb shift of the two-level atom
\cite{31}. Here we restrict our attention to the weak coupling
regime, where the Markov approximation \cite{31,32} applies. That
is to say, we may replace $ c(t')$ in the integrand in (\ref{d47})
by
\begin{equation}\label{d52}
c(t')=c(t)e^{\imath\omega_0(t-t')},
\end{equation}
 and approximate the time integral in (\ref{d47}) in very large times by
\begin{equation}\label{d53}
\int_0^t dt' e^{i(\omega_0-\omega)(t-t')}\approx
\left[iP\frac{1}{\omega_0-\omega}+\pi\delta(\omega_0-\omega)\right],
\end{equation}
where $ P $ denotes the principal cauchy value. After some simple
algebra and using  the Kramers-Kronig relations for the diadic $
\tilde{G}$ we deduce
\begin{equation}\label{d53}
\dot{c}(t)=-\imath\omega_0c(t)-(\Gamma_0+\Gamma+\imath\Delta_0+\imath\Delta)c(t),
\end{equation}
where
\begin{eqnarray}\label{d52}
 &&\Gamma=-\frac{1}{\hbar}
\left[\vec{r}_{12}\cdot I m
\tilde{G}(\vec{R},\vec{R},\omega_0+\imath0^+
)\cdot\vec{r}_{12}\right],\nonumber\\
&& \Delta=\frac{1}{\hbar} \left[\vec{r}_{12}\cdot Re
\tilde{G}(\vec{R},\vec{R},\omega_0+\imath0^+
)\cdot\vec{r}_{12}\right],
\end{eqnarray}
 are  decay constant and the level shift due
to the presence of the magnetodielectric, respectively.

There are many cases that the medium is homogeneous inside the
region $\Omega$ and
$W_{\vec{n}\lambda}^{\vec{n}\lambda}(\imath\rho)$ in the
denominator of $\tilde{G}$ in (\ref{d49}) is independent of
polarization label $\lambda$ and vector label $\vec{n}$. For
example for the box $ 0<x<L_1$, $0<y<L_2$, $0<z<\frac{L_3}{2}$
filled up uniformly with a homogeneous magnetodielectric, from the
definition (\ref{d43}), we have
\begin{equation}\label{d53}
W_{\vec{n}\lambda}^{\vec{n}\lambda}(\imath\rho)\equiv
W(\imath\rho)=\frac{1}{2}\left(Z_e(\imath\rho)+Z_m(\imath\rho)\right).
\end{equation}
Generally speaking for a piece wisely homogeneous medium with $ N
$ homogeneous pieces in volumes $\Omega_1$, $\Omega_2$,$\cdots$,
$\Omega_N $, applying the definitions (\ref{d1.1}) and (
\ref{d4.735}) in (\ref{d43}), we note that
$W_{\vec{n}\lambda}^{\vec{n}\lambda}( \imath\rho)$ in denomonator
of $ \tilde{G}$ can be approximated by
\begin{equation}\label{d55}
W_{\vec{n}\lambda}^{\vec{n}\lambda}(\imath\rho)\equiv
W(\imath\rho)\simeq
\sum_{i=1}^N\frac{\Omega_i}{V}\left(Z_e^{(i)}(\imath\rho)+Z_m^{(i)}(\imath\rho)\right),
\end{equation}
 where $ Z_e^{(i)} $ and $ Z_e^{(i)}$ denote the quantities $ Z_e $ and $Z_m $ for
 the ith piece respectively. Now we can substitute $
W_{\vec{n}\lambda}^{\vec{n'}\lambda'}(\imath\rho)$ from
(\ref{d43}) in the numerator of $\tilde{G}$ given by (\ref{d49})
and do the summations over $\lambda,\lambda'$ by using the
completeness relations
\begin{eqnarray}\label{d56}
&&\sum_{\lambda=1}^2u^i_{\vec{n}\lambda}(\vec{R})u^j_{\vec{n}\lambda}(\vec{r})=(\delta_{ij}-
\hat{k}^i_{\vec{n}}\hat{k}^j_{\vec{n}})
f_i(\vec{n},\vec{R})f_j(\vec{n},\vec{r}),\nonumber\\
&&\sum_{\lambda=1}^2u^i_{\vec{n}\lambda}(\vec{R})s^j_{\vec{n}\lambda}(\vec{r})=\sum_{\mu=1}^3\varepsilon_{j\mu
i}\hat{k}_{\vec{n}\mu}f_i(\vec{n},\vec{R})g_j(\vec{n},\vec{r}).
\end{eqnarray}
Now in the box frame we have
\begin{eqnarray}\label{d57}
&&\tilde{G}_{\alpha\beta}(\vec{R},\vec{R},\imath\rho)=\frac{2e^2}{\varepsilon_0}\sum_{i=1}^N\int_{\Omega_i}
d^3r\sum_{\gamma=1}^3 \left\{Z_e^{(i)}(\imath\rho)\eta^e_{\alpha
\gamma}\zeta^e_{\beta\gamma}+Z_m^{(i)}(\imath\rho)
\eta^m_{\alpha\gamma}\zeta^m_{\beta\gamma}\right\},\nonumber\\
&&
\end{eqnarray}
where
\begin{eqnarray}\label{d58}
&&\eta_{\alpha\gamma}^e(\vec{R},\vec{r},\imath\rho)=
\frac{4}{V}\sum_{\vec{n}}\frac{\omega_{\vec{n}}}
{\imath\rho-\omega_{\vec{n}}\left[1+W(\imath\rho)\right]}
\left(\delta_{\alpha\gamma}-\hat{k}^\alpha_{\vec{n}}\hat{k}^\gamma_{\vec{n}}\right)f_\alpha(\vec{n},\vec{R})
f_\gamma(\vec{n},\vec{r}),\nonumber\\
&&\zeta_{\beta\gamma}^e(\vec{R},\vec{r},\imath\rho)=\frac{4}{V}\sum_{\vec{n}}\frac{\omega_{\vec{n}}}
{\imath\rho-\omega_{\vec{n}}}
\left(\delta_{\beta\gamma}-\hat{k}^\beta_{\vec{n}}\hat{k}^\gamma_{\vec{n}}\right)f_\beta(\vec{n},\vec{R})
f_\gamma(\vec{n},\vec{r}),\nonumber\\
&&\eta_{\alpha\gamma}^m(\vec{R},\vec{r},\imath\rho)=\frac{4}{V}\sum_{\vec{n}}\sum_{\mu=1}^3\frac{\omega_{\vec{n}}}
{\imath\rho-\omega_{\vec{n}}\left[1+W(\imath\rho)\right]}
\varepsilon_{\gamma\mu
\alpha}\hat{k}^\mu_{\vec{n}}f_\alpha(\vec{n},\vec{R})g_\gamma(\vec{n},\vec{r}),\nonumber\\
&&\eta_{\beta\gamma}^m(\vec{R},\vec{r},\imath\rho)=\frac{4}{V}\sum_{\vec{n}}\sum_{\mu=1}^3\frac{\omega_{\vec{n}}}
{\imath\rho-\omega_{\vec{n}}} \varepsilon_{\gamma\mu
\beta}\hat{k}^\mu_{\vec{n}}f_\beta(\vec{n},\vec{R})g_\gamma(\vec{n},\vec{r})
\end{eqnarray}
It is remarkable to note that since we have assumed that the
center of the atom is localized in a free region, then in the
argument of the tensors $\eta^e$, $\zeta^e$, $\eta^m$ and
$\zeta^m$ in (\ref{d58}), we have $\vec{R}\neq\vec{r}$ and
therefore the recent series are all convergent. Inspection of the
definitions (\ref{d44}) shows that the real and imaginary parts of
$Z_e(\omega+\imath 0^+)$ and $Z_m(\omega+\imath 0^+ )$ for real
frequency $\omega $ are
\begin{eqnarray}\label{d59}
&& Re[Z_e(\omega+\imath 0^+)]=\frac{1}{2\pi}P\int_0^\infty
d\omega'\frac{\chi_{ei}(\omega')}{\omega-\omega'}\equiv\alpha_e(\omega),\nonumber\\
&&I m[Z_e(\omega+\imath
0^+)]=-\frac{1}{2}\chi_{ei}(\omega)\vartheta(\omega)\equiv\gamma_e(\omega),\nonumber\\
&& Re[Z_m(\omega+\imath 0^+)]=\frac{1}{2\pi}P\int_0^\infty
d\omega'\frac{\chi_{mi}(\omega'}{\omega-\omega'}\equiv\alpha_m(\omega),\nonumber\\
&&I m[Z_m(\omega+\imath
0^+)]=-\frac{1}{2}\chi_{mi}(\omega)\vartheta(\omega)\equiv\gamma_m(\omega),
\end{eqnarray}
where $\vartheta(\omega) $ is the step function and $P$ denotes
the Causchy principal value. Finally using (\ref{d52}) ,
(\ref{d57}), (\ref{d58}), (\ref{d59}) and
\begin{equation}\label{d4}
\frac{1}{\omega+\imath0^+-\omega'}=P\frac{1}{\omega-\omega'}-\imath\pi\delta(\omega-\omega'),
\end{equation}
we find the spontaneous emission and the shift frequency of a two
level atom due to the presence of a magnetodielectric. From
(\ref{d52}) and (\ref{d57}) it is clear that the spontaneous
emission and the lamb shift depend on the (i) position of the
atom, (ii) orientation of the dipole of the atom,(iii) size and
sort of the medium, (iv) relative position of the medium respect
to the center of the atom, (v) the volume and the shape of the
cavity. For a a non rectangular cavity we must use a different set
of eigenvectors and therefore the spontaneous emission depends
also on the shape of the cavity.
\section{Conclusion}
The scheme introduced by the present authors to quantize the
electromagnetic field in a magnetodielectric is generalized to the
case where some external charges are present in the medium. The
spontaneous emission of a two-level atom embedded in a
magnetodielectric is calculated as an application of the model.

\end{document}